\documentclass{article}

    \usepackage[final]{tackling_climate_workshop_style}

\usepackage{color}
\usepackage[utf8]{inputenc} 
\usepackage[T1]{fontenc}    
\usepackage{hyperref}       
\usepackage{url}            
\usepackage{booktabs}       
\usepackage{caption}
\usepackage{threeparttable}
\usepackage{amsfonts}       
\usepackage{nicefrac}       
\usepackage{microtype}      
\usepackage{graphicx}
\usepackage{float}          
\usepackage{doi}

\title{Explainable Multi-Agent Recommendation System for
Energy-Efficient Decision Support
in Smart Homes
}

\author{\href{https://orcid.org/0000-0003-3506-4744}{\includegraphics[scale=0.06]{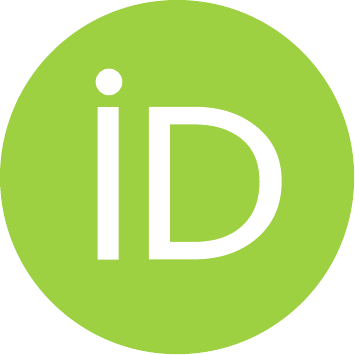}
Alona Zharova\thanks{Corresponding author.} } \\
	Humboldt-Universität zu Berlin\\
	Berlin, Germany \\
	\texttt{alona.zharova@hu-berlin.de} \\
	\And
	Annika Boer \\
	Humboldt-Universität zu Berlin\\
	Berlin, Germany \\
	\texttt{boeranni@hu-berlin.de} \\
	\And
	Julia Knoblauch \\
	Humboldt-Universität zu Berlin\\
	Berlin, Germany \\
	\texttt{julia.knoblauch@hu-berlin.de} \\
	\And
	Kai Ingo Schewina \\
	Humboldt-Universität zu Berlin\\
	Berlin, Germany \\
	\texttt{schewink@hu-berlin.de} \\
	\And
	Jana Vihs \\
	Humboldt-Universität zu Berlin\\
	Berlin, Germany \\
	\texttt{vihsjana@hu-berlin.de} \\
}

\begin{document}

\maketitle

\begin{abstract}
    Understandable and persuasive recommendations support the electricity consumers’ behavioral change to tackle the energy efficiency problem.
    This paper proposes an explainable multi-agent recommendation system for load shifting for the household appliances.
    First, we provide agents with enhanced predictive capacity by including weather data, applying state-of-the-art models and tuning the hyperparameters. 
    Second, we suggest an Explainability Agent providing
    transparent 
    recommendations.
    Third, we discuss the potential impact of the suggested approach.
\end{abstract}


\section{Introduction}
Europe faces a double urgency to increase energy efficiency: on the one hand, caused by the war in Ukraine, on the other hand, due to the continuous rise in electricity consumption [1].
Tackling the energy efficiency problem through consumers' behavioral change is an obvious, however challenging solution. 
People often need a guidance, and sometimes a soft nudge to put the intentions into actions [2], 
for instance, to change the timing of appliances usage.
Recommender systems can suggest energy-efficient actions to facilitate such behavioral change. 
To increase the trust in the recommendation system,  and, thus, the acceptance rate of recommendations,
users need to understand why and how the model makes its predictions [3], [4]. 
Thus, the recommendation system should be explainable. 

The existing research on explainability in recommender systems for energy-efficient smart homes is very scarce [5]. [6] provide a thorough literature review on explainability in recommender systems for other application domains. However, most existing approaches are not applicable to the smart home area because of the missing data structures. [7] design an explainable context-aware recommendation system for a smart home ecosystem. They show that displaying the advantages and the reasoning behind recommendations lead to a 19\% increase in acceptance rate. 
To our best knowledge, the issue of explainability in multi-agent recommendation systems for energy-efficient smart homes has not been studied yet.

Our contributions are twofold. First, we suggest an explainable multi-agent recommendation system for energy efficiency in private households. In particular, we extend a novel multi-agent approach of [8] by designing an Explainability Agent that provides explainable recommendations for optimal appliance scheduling in a textual and visual manner. Second, we enhance the predictive capacity of other agents by including weather data and applying state-of-the-art models. 
We also provide an overview of predictive and explainability performance.
We provide a comprehensive tutorial in Jupyter Notebook 
in GitHub \footnote[1]
{\url{{https://github.com/Humboldt-WI/Explainable_multi-agent_RecSys}}}
for all the steps described in this paper and beyond. 



\section{Explainable multi-agent recommendation system}

[8] introduce a utility-based context-aware multi-agent recommendation system that provides load shifting recommendations for  household devices for the next 24 hours. Their system includes six agents: Price Agent (prepares external hourly electricity prices), 
Preparation Agent (prepares data for the other agents), 
Availability Agent (predicts the hourly user availability for the next 24 hours), 
Usage Agent (calculates the devices' usage probabilities for the prediction day),
Load Agent (extracts the typical devices' loads), and Recommendation Agent (collects the inputs from the other agents and provides recommendations). 
The multi-agent architecture of [8] is flexible and can be easily integrated into existing smart home systems. However, the cost of the simplicity of the approach (i.e., they use Logistic Regression for the availability and usage predictions) is a relatively low prediction accuracy.

\begin{figure}[H]
  \centering
	\includegraphics[width=1\textwidth]{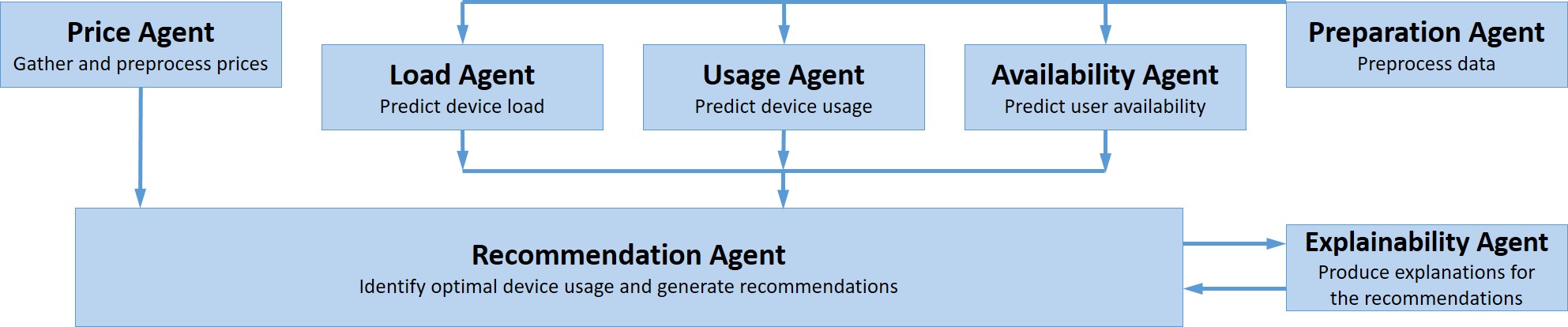}
  \caption{Architecture of the explainable multi-agent recommendation system.}
  \label{fig:structure}
\end{figure}

We address the limitations in [8] by 
enhancing the performance of the Availability and the Usage Agents. In particular, we apply the K-Nearest-Neighbours (KNN), extreme gradient boosting (XGBoost), adaptive boosting (AdaBoost), and Random Forest to predict the availability and usage probabilities. 
Furthermore, we use Logistic Regression (Logit) 
and Explainable Boosting Machines (EBM, see [9])
as inherently explainable models.
We propose including the Explainability Agent in the system (see Figure \ref{fig:structure}).
The explainability models are divided into local and global, depending on their capability to explain a particular instance or the entire model. 
Since we want to help the user understand a single recommendation, we focus on local approaches.
In particular, we apply post-model approaches LIME (local, interpretable, model-agnostic explanation; [10]) and SHAP (Shapley additive explanations; [11]) as model-agnostic tools that can explain the predictions of the chosen classifiers.

To create an explanation for a recommendation, the Explainability Agent extracts feature importance from the explainability models
. 
We design the explanation 
to include two parts: (i) Usage explanation - which features lead to the specific device usage prediction for the day? and (ii) Availability explanation - which features drive the user availability prediction for the hour? We do not include an explanation for the Load Agent since we do not consider the extracted typical load profile of every shiftable device as informative to the users.
As a result, the Recommendation Agent recommends the cheapest starting hour within the hours of user availability for the shiftable devices that are likely to be used on the prediction day with an explanation in a text and visual form. The system provides no recommendations if the predictions for the availability and usage probabilities are below the thresholds.


\section{Experimental design}

We use the REFIT data set on electrical load measurements, provided in [12]. It contains the energy consumption of nine devices in Watts used in 20 households in the United Kingdom as well as aggregate energy consumption in each household over the period 2013 to 2015.
Additionally, we utilize the weather data from the meteostat, provided in [13]. In particular, we use the following features: dew point, relative humidity, air temperature, average wind direction, and average wind speed.
The missing values are imputed using the KNN algorithm.

We apply an exhaustive grid search over the algorithms mentioned above, excluding the EBM. The latter is computationally expensive. Overall, 87 parameter combinations are tested twice (with and without weather data) to quantify the benefit of including additional data and pursuing a fusion-based approach. 
We use a KernelExplainer for explaining Logit, AdaBoost, KNN and Random Forest. 
For XGBoost, we use the fastest version of the TreeExplainer - interventional feature perturbation [14]. 

To evaluate the performance of the multi-agent model we apply multiple metrics depending on the task. The Usage and the Availability Agents perform a classification task, and therefore, we evaluate their area-under-the-ROC-curve (AUC). 
The Load Agent is evaluated by investigating the mean squared error (MSE) of the predicted load profile to real usages. 
Next, we use three metrics within the Explainability Agent to reflect how well the explainability approaches work in giving accurate local explanations [15]: accuracy, fidelity, and efficiency. Accuracy shows how well the explainable model predicts unseen instances compared to the real outcome. 
Fidelity determines 
how close the prediction from the explainable model is to the black-box model's prediction. 
Additionally, we calculate the mean absolute explainability error (MAEE) for every approach. 
The efficiency metric describes the algorithmic complexity of the explainability approaches when calculating the local explanations. 
Lastly, we measure the time that each method needs to calculate all the local explanations for a day and average the values for all calculations.

\section{Results}

Using the prediction methods mentioned above, we observe consistent results for the Availability Agent. It benefits less from complex models and the inclusion of weather data than the Usage Agent. 
Random Forest model learns at most from additional data. The range between best and worst performing models is 0.023 AUC points regardless of weather inclusion, which indicates robustness towards algorithmic choice as well as weather features.
Contrarily, the Usage Agent benefits greatly from the weather data. While without the additional data, most models perform around 0.7 in AUC. The inclusion allows for substantial increases toward the performance of 0.93. Especially more complex models (i.e., Random Forest, XGBoost, Adaboost) profit from the inclusion and outperform the approach of [8]. We conclude that it is not sufficient to either include more complex models or weather data, but the combination of the two leads to the substantial performance improvements.
For more details, please refer to the Appendix.

Since Random Forest and XGBoost perform similar in their respective best setup, we further analyze the AUC over ten households to investigate their stability. 
We notice that Random Forest slightly but consistently outperforms XGBoost. On average, XGBoost achieves an AUC of 0.884 over all prediction tasks and devices. While the Random Forest achieves an AUC of 0.891. The EBM performs similarly to the alternatives for the Availability Agent. 
We decide to use further Random Forest with the tuned parameters derived from the grid search.

\subsection{Explainability}

We examine the results of the explainability evaluation by using the predictions from LIME and SHAP.
SHAP generally performs better according to the higher accuracy and fidelity compared to LIME. Most strikingly, since the forecasts of the prediction algorithms and SHAP are so similar, the fidelity is almost perfect. In other words, SHAP works very well for mimicking the local behavior of the prediction model. 
The accuracy of LIME for the Usage Agent is  not so high. The possible reason is that the predictions, e.g., for AdaBoost, are very close to the chosen cutoff of 0.5. 
Although LIME is only off on average about 0.0311 AUC points, it sometimes exceeds the cutoff assigning it to the wrong class. For the Availability Agent, the worse calibration of LIME does not take too much influence since the values from the prediction are more extreme.
Furthermore, for most cases SHAP creates predictions faster than LIME. Especially, LIME is not able to handle XGBoost very well while the TreeExplainer of SHAP used for XGBoost provides a very good runtime. 

Since there is no significant difference between the fidelity of SHAP across the models, we apply SHAP with one of the most accurate algorithms while also minimizing runtime. 
XGBoost and Random Forest perform similarly regarding predictive performance for the Availability Agent, therefore, we also run an evaluation with the TreeExplainer for Random Forest. The resulting runtimes (Availability: 0.558, Usage: 0.0357) are also low, while there is little change with regard to the metrics. Thus, we select Random Forest and SHAP with the TreeExplainer as our final model.

\subsection{Explainable recommendation}

\begin{figure}[H]
  \centering
	\includegraphics[width=1\textwidth]{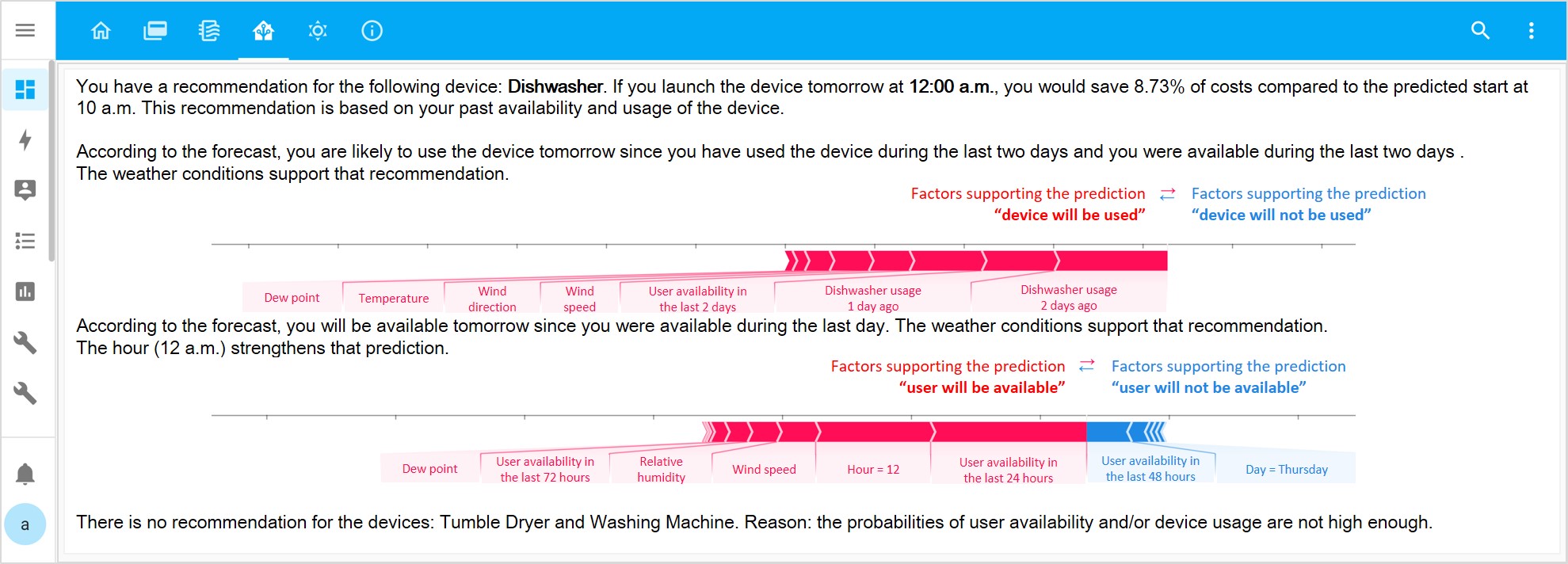}
  \caption{Exemplary explainable recommendation.}
  \label{fig:recom}
\end{figure}

The recommendation for the next 24 hours is provided once a day at the same time specified by the user.
To create an explanation for the recommendation, we create a feature importance ranking using Shap Values.
We provide two different explanations for the Availability and the Usage Agents. 
We separate the features in two categories for each of the agents: 
features based on weather data 
and non-weather features.
The Recommendation Agent embeds the two most important features of each group into an explanation sentence. The usage explanation is provided for each device (if its usage is recommended) since their predictions differ. Additionally, we adapt the plots provided by SHAP to inform the user about the specific impact of the features. 
We only display the most important features to shorten the explanation. We show an exemplary explainable recommendation in Figure \ref{fig:recom}.

\section{Potential impact}

Our medium-term goal is to integrate the explainable recommendation system with the existing smart home platform used in German residential houses.
Seeing the recommendations daily will increase the awareness of residents of their energy consumption and will encourage more climate-related actions (supporting SDG 13).
The shifted load will facilitate energy efficiency in the grid (SDG 7), foster energy innovation towards sustainable development (SDG 9), reduce the environmental impact, and stronger the households' sustainability making them inclusive, safe, and resilient (SDG 11).

\section{Conclusions}

Our results show a substantial improvement in the performance of the multi-agent recommendation system while at the same time opening up the “black box” of recommendations. 
Generating load shifting recommendations for household appliances as explainable increases the transparency and  trustworthiness of the system. As a result, the understandable and persuasive recommendations facilitate the consumers' behavioral change in tackling the energy efficiency problem.




\section*{Acknowledgments}

We would like to thank Lucas Pereira for his great mentorship within the CCAI program while preparing this paper for the submission.
We wish to express our sincere gratitude and
warm appreciation to Stefan Lessmann for many insightful discussions on the topic. 
We are deeply thankful to all participants of the Information Systems Research Seminar at the Humboldt-Universität zu Berlin for the valuable feedback.

\section*{References}

\small

[1] European Commission. (2022). \textit{REPowerEU: A plan to rapidly reduce dependence on Russian fossil fuels and fast forward the green transition.} 18 May 2022. Retrieved from \url{https://ec.europa.eu/commission/presscorner/detail/en/IP_22_3131}. 

[2] Frederiks, E.R., Stenner, K., \& Hobman, E. V. (2015). Household energy use: Applying behavioural economics to understand consumer decision-making and behaviour. \textit{Renewable and Sustainable Energy Reviews, 41}, 1385–1394. 

[3] Luo, F., Ranzi, G., Kong, W., Liang, G., \& Dong, Z.Y. (2021). Personalized Residential Energy Usage Recommendation System Based on Load Monitoring and Collaborative Filtering. \textit{IEEE Transactions on Industrial Informatics, 17}(2), 1253-1262.

[4] Sayed, A., Himeur, Y., Alsalemi, A., Bensaali, F., \& Amira, A. (2022). Intelligent Edge-Based Recommender System for Internet of Energy Applications.\textit{ IEEE Systems Journal, 16}(3), 5001–5010.

[5] Himeur, Y., Alsalemi, A., Al-Kababji, A., Bensaali, F., Amira, A., Sardianos, C., Dimitrakopoulos, G., \& Varlamis, I. (2021). A survey of recommender systems for energy efficiency in buildings: Principles, challenges and prospects. \textit{Information fusion, 72}(2021), 1-21.

[6] Zhang, Y., \& Chen, X. (2020). Explainable recommendation: A survey and new perspectives. {\it Foundations and trends in information retrieval, 14}(1), 1-101.

[7] Sardianos, C., Varlamis, I., Chronis, C., Dimitrakopoulos, G., Alsalemi, A., Himeur, Y., Bensaali, F., \& Amira, A. (2021). The emergence of explainability of intelligent systems: Delivering explainable and personalized recommendations for energy efficiency. \textit{International Journal of Intelligent Systems, 36}(2), 656–680.

[8] Riabchuk, V., Hagel, L., Germaine, F. \& Zharova, A. (2022). Utility-Based Context-Aware Multi-Agent Recommendation System for Energy Efficiency in Residential Buildings. \textit{arXiv:2205.02704}. 

[9] Nori, H., Jenkins, S., Koch, P., \& Caruana, R. (2019). Interpretml: A unified framework for machine learning interpretability. \textit{arXiv:1909.09223.}

[10] Ribeiro, M.T., Singh, S., \& Guestrin, C. (2016). Why Should I Trust You? Explaining the predictions of any classifier. \textit{KDD'16: Proceedings of the 22nd ACM SIGKDD International Conference on Knowledge Discovery and Data Mining.} pp. 1135–1144.

[11] Lundberg, S.M., \& Lee, S.I. (2017). A unified approach to interpreting model predictions. \textit{NIPS'17: Proceedings of the 31st International Conference on Neural Information Processing Systems}. pp. 4768-4777.

[12] Murray, D., Stankovic, L. \& Stankovic, V. (2017). An electrical load measurements dataset of united kingdom households from a two-year
longitudinal study. \textit{Scientific Data, 4}(1), 160122.

[13] Meteostat. (2022). GitHub repository  meteostat/meteostat-python.  
Access and analyze historical weather and climate data with Python. Retrieved from \url{https://github.com/meteostat/meteostat-python}. 

[14] Lundberg, S.M., Erion, G., Chen, H., Degrave, A., Prutkin, J.M., Nair, B., Katz, R., Himmelfarb, J., Bansal, N., \& Lee, S.-I. (2020). From local explanations to global understanding with explainable AI for trees. \textit{Nature Machine Intelligence, 2}(1), 56–67. 

[15] Carvalho, D.V., Pereira, E.M., \& Cardoso, J.S. (2019). Machine learning interpretability: A survey on methods and metrics. \textit{Electronics, 8}(8), 832.

[16] Statista Research Department. (2021). \textit{Prognose zur Anzahl der Smart Home Haushalte nach Segmenten in Deutschland für die Jahre 2017 bis 2025 (in Millionen).} [Infographic]. Statista. Retrieved from \url{https://de.statista.com/prognosen/801573/anzahl-der-smart-home-haushalte-nach-segmenten-in-deutschland}. 

[17] Statistisches Bundesamt [Destatis]. [2020]. \textit{Entwicklung der Privathaushalte bis 2040.} Retrieved from \url{https://www.destatis.de/DE/Themen/Gesellschaft-Umwelt/Bevoelkerung/Haushalte-Familien/Publikationen/Downloads-Haushalte/entwicklung-privathaushalte-5124001209004.pdf?__blob=publicationFile}. 

[18] Statista Research Department (2022). \textit{Jährlicher Stromverbrauch eines privaten Haushaltes in Deutschland in den Jahren 1991 bis 2021 (in Kilowattstunden)}. [Infographic]. Statista. Retrieved from \url{https://de.statista.com/statistik/daten/studie/245790/umfrage/stromverbrauch-eines-privathaushalts-in-deutschland/}. 

[19] Icha, P., Lauf, T. \& Kuhs, G. (2022). Entwicklung der spezifischen Kohlendioxid - Emissionen des deutschen Strommix in den Jahren 1990 - 2020. Umweltbundesamt. {\it Climate Change, 15}(2022). Retrieved from \url{https://www.umweltbundesamt.de/sites/default/files/medien/1410/publikationen/2022-04-13_cc_15-2022_strommix_2022_fin_bf.pdf}. 


\newpage

\section*{Appendix}

\begin{table}[H]
  \caption{Performance evaluation results (in AUC) for the Availability and the Usage Agents with tuned hyperparameters including weather data}
  \centering
  \begin{tabular}{lrrrr}
    \toprule
    Model & Availability & Usage TD & Usage WM & Usage DW \\
    \midrule
    Logit &	0.836 &	0.728 &	0.703 &	0.684 \\
    Random Forest &	\textbf{0.859} &	\textbf{0.928} &	\textbf{0.932} &	\textbf{0.929} \\
    AdaBoost &	0.850 &	0.779 &	0.777 &	0.788 \\
    KNN &	0.830 &	0.760 &	0.758 &	0.785 \\
    XGBoost &	0.852 &	\underline{0.922} &	\underline{0.925} &	\underline{0.922} \\
    EBM &	\underline{0.854} &	0.803 &	0.822 &	0.842 \\
    \bottomrule
  \end{tabular}
    \begin{tablenotes}
      \small
      \item Abbreviations: TD - Tumble Dryer, WM - Washing Machine, DW - Dishwasher.
      \item The best performing model is in bold, the second-best - is underlined.
    \end{tablenotes}
  \label{tab:A1}
\end{table}

\begin{table}[H]
  \caption{Performance evaluation results (in AUC) for the Availability and the Usage Agents with tuned hyperparameters excluding weather data}
  \centering
  \begin{tabular}{lrrrr}
    \toprule
    Model & Availability & Usage TD & Usage WM & Usage DW \\
    \midrule
    Logit &	0.838 &	0.702 &	0.689 &	0.676 \\
    Random Forest &	0.835 &	\underline{0.713} &	0.683 &	\underline{0.715} \\
    AdaBoost &	\textbf{0.849} &	0.710 &	\underline{0.701} &	0.680 \\
    KNN &	0.826 &	\textbf{0.738} &	\textbf{0.718} &	\textbf{0.779} \\
    XGBoost &	0.845 &	0.712 &	0.682 &	0.710 \\
    EBM &	\underline{0.846} &	0.707 &	0.683 &	0.700 \\
    \bottomrule
  \end{tabular}
    \begin{tablenotes}
      \small
      \item Abbreviations: TD - Tumble Dryer, WM - Washing Machine, DW - Dishwasher.
      \item The best performing model is in bold, the second-best - is underlined.
    \end{tablenotes}
  \label{tab:A2}
\end{table}

\begin{table}[H]
  \caption{Performance evaluation results (in AUC) for Random Forest for the Availability and the Usage Agents for the ten households}
  \centering
  \begin{tabular}{lrrrrrr}
    \toprule
    Household & Availability & Usage TD & Usage WM & Usage DW & Usage WM2 & Usage WD \\
    \midrule
    h1 &	0.795 &	0.949 &	0.923 &	0.942  && \\		
    h2 &	0.827 	&&	0.892 &	0.932 	&& \\	
    h3 &	0.859 &	0.927 &	0.934 &	0.927 	&& \\	
    h4 &	0.860 	&&	0.929 	&&	0.871 	& \\
    h5 &	0.816 &	0.910 &&&& \\				
    h6 &	0.826 	&&	0.914 &	0.948 	&& \\	
    h7 &	0.881 &	0.950 &	0.945 &	0.945 & & \\		
    h8 &	0.675 	&&	0.921 	&&& \\		
    h9 &	0.765 	&&	0.926 &	0.944 &&		0.920 \\
    h10 &	0.820 	&&	0.950 	&& \\		
    \bottomrule
  \end{tabular}
    \begin{tablenotes}
      \small
      \item Abbreviations: TD - Tumble Dryer, WM - Washing Machine, DW - Dishwasher, WM2 - 2nd Washing Machine, WD - Washdryer.
    \end{tablenotes}
  \label{tab:A3}
\end{table}

\begin{table}[H]
  \caption{Performance evaluation results (in AUC) for XGBoost for the Availability and the Usage Agents for the ten households}
  \centering
  \begin{tabular}{lrrrrrr}
    \toprule
    Household & Availability & Usage TD & Usage WM & Usage DW & Usage WM2 & Usage WD \\
    \midrule
    h1 &	0.780 &	0.927 &	0.910 &	0.936 	&& \\	
    h2 &	0.817 	&&	0.890 &	0.928 	&& \\	
    h3 &	0.852 &	0.922 &	0.925 &	0.922 	& & \\	
    h4 &	0.855 	&&	0.917 	&&	0.860 	& \\
    h5 &	0.817 &	0.905 &&& \\				
    h6 &	0.825 &	&	0.905 &	0.945 	& & \\	
    h7 &	0.884 &	0.943 &	0.937 &	0.930 & & \\		
    h8 &	0.671 	&&	0.918 	&&& \\		
    h9 &	0.764 	&&	0.914 &	0.936 	&&	0.916 \\
    h10 &	0.800 	&&	0.942 	&&& \\	
    \bottomrule
  \end{tabular}
    \begin{tablenotes}
      \item Abbreviations: TD - Tumble Dryer, WM - Washing Machine, DW - Dishwasher, WM2 - 2nd Washing Machine, WD - Washdryer.
    \end{tablenotes}
  \label{tab:A4}
\end{table}

\begin{table}[H]
    \caption {Performance evaluation results for Logistic Regression for the ten households from [8]}
    \centering
    \begin{tabular}{llrrrrrr}
        \toprule
        {} & Availability AUC & \multicolumn{3}{l}{Usage AUC} & \multicolumn{3}{l}{Load MSE} \\
         Devices &              &           &       &       &           &          &          \\
        Households &                 &         \multicolumn{1}{l}{0} &     \multicolumn{1}{l}{1} &     \multicolumn{1}{l}{2} &         \multicolumn{1}{l}{0} &        \multicolumn{1}{l}{1} &        \multicolumn{1}{l}{2} \\
        \midrule
        1         &         0.72 &      0.52 &  0.45 &  0.49 &   1224.45 &   511.49 &  3021.88 \\
        2         &         0.77 &      0.60 &   0.80 &     - &    320.81 &  14578.80 &        - \\
        3         &         0.80 &      0.65 &  0.65 &  0.65 &  24907.50 &   890.33 &   314.84 \\
        4         &         0.82 &      0.67 &  0.52 &     - &      0.80 &  1793.77 &        - \\
        5         &         0.76 &      0.68 &     - &     - &  37982.21 &        - &        - \\
        6         &         0.77 &      0.53 &  0.66 &     - &    424.78 &  6032.65 &        - \\
        7         &         0.83 &      0.79 &   0.80 &  0.81 &  43673.35 &   336.83 &  6845.26 \\
        8         &         0.62 &      0.77 &     - &     - &      0.35 &        - &        - \\
        9         &         0.67 &      0.60 &  0.57 &  0.78 &  30711.30 &    14.36 &  5480.13 \\
        10        &         0.77 &      0.74 &     - &     - &    647.57 &        - &        - \\
        \bottomrule
    \end{tabular}
    \begin{tablenotes}
      \small
      \item The table is reproduced here from [8] to enable the direct comparison.
      \item Legend for mapping the shiftable devices 0 - 2 is provided in Table \ref{tab:A6}.
    \end{tablenotes}
    \label{tab:A5}
\end{table}

\begin{table}[H]
    \caption{\label{tab:device-index-legend}Legend for mapping the shiftable devices to an integer index from [8]}
    \centering
    \begin{tabular}{llll}
        \toprule
                  &                     &                 Devices &            \\
        Household &                    0  &                   1  &          2  \\
        \midrule
        1         &         Tumble Dryer &      Washing Machine &  Dishwasher \\
        2         &      Washing Machine &           Dishwasher &           - \\
        3         &         Tumble Dryer &      Washing Machine &  Dishwasher \\
        4         &  Washing Machine (1) &  Washing Machine (2) &           - \\
        5         &         Tumble Dryer &                    - &           - \\
        6         &      Washing Machine &           Dishwasher &           - \\
        7         &         Tumble Dryer &      Washing Machine &  Dishwasher \\
        8         &      Washing Machine &                    - &           - \\
        9         &         Washer Dryer &      Washing Machine &  Dishwasher \\
        10        &      Washing Machine &                    - &           - \\
        \bottomrule
    \end{tabular}
    \label{tab:A6}
\end{table}

\begin{table}[H]
  \caption{Explainability evaluation results for LIME and SHAP for the Availability Agent for the ten households}
  \label{tab:A7}
  \centering
  \begin{tabular}{lrrrrrrrr}
    \toprule
    \multicolumn{1}{c}{} & \multicolumn{4}{c}{LIME} & \multicolumn{4}{c}{SHAP} \\
    \cmidrule(r){2-9}
    Model &	Accuracy &	Fidelity &	MAEE	& Duration	& Accuracy &	Fidelity &	MAEE	& Duration \\
    \midrule
    Logit &	0.8371 &	0.9890 &	0.0265 &	73.748 &	0.8362 &	1 &	0 &	2.147 \\
    KNN &	0.8295 &	0.9747 &	0.0194 &	79.695 &	0.8386 &	1 &	0 &	20.759 \\
    Adaboost &	0.8369 &	0.9349 &	0.0035 &	50.975 &	0.8479 &	1 &	0 &	52.103 \\
    Random F. &	0.8411 &	0.9761 &	0.0313 &	107.363 &	0.8451 &	1 &	0 &	64.853 \\
    XGBoost &	0.8378 &	0.7971 &	0.1848 &	310.551 &	0.8366 &	1 &	0 &	0.717 \\
    \bottomrule
  \end{tabular}
    \begin{tablenotes}
      \item The values for MAE are not displayed as they are too small. The duration is only comparable within the model since the evaluation was run on different machines with different background tasks.
    \end{tablenotes}
\end{table}

\begin{table}[H]
  \caption{Explainability evaluation results for LIME and SHAP for the Usage Agent for the ten households}
  \label{tab:A7}
  \centering
  \begin{tabular}{lrrrrrrrr}
    \toprule
    \multicolumn{1}{c}{} & \multicolumn{4}{c}{LIME} & \multicolumn{4}{c}{SHAP} \\
    \cmidrule(r){2-9}
    Model &	Accuracy &	Fidelity &	MAEE	& Duration	& Accuracy &	Fidelity &	MAEE	& Duration \\
    \midrule
    Logit &	0.6761 &	0.8215 &	0.1454 &	2.582 &	0.7049 &	1 &	0 &	0.084 \\
    KNN	 & 0.6928 &	0.8101 &	0.1858 &	2.864 &	0.7611 &	1 &	0 &	0.852 \\
    Adaboost &	0.5757 &	0.5934 &	0.0311 &	1.859 &	0.8254 &	1 &	0 &	0.813 \\
    Random F. &	0.7101 &	0.7988 &	0.1203 &	3.682 &	0.8726 &	1 &	0 &	0.957 \\
    XGBoost &	0.6763 &	0.6516 &	0.2705 &	10.222 &	0.9353 &	1 &	0 &	0.230 \\
    \bottomrule
  \end{tabular}
    \begin{tablenotes}
      \item The values for MAE are not displayed as they are too small. The duration is only comparable within the model since the evaluation was run on different machines with different background tasks.
    \end{tablenotes}
\end{table}

\end{document}